\documentclass[twocolumn,prb,showpacs]{revtex4}
\usepackage{graphicx}
\usepackage{dcolumn}
\usepackage{bm}
\usepackage[cp1251]{inputenc}
\usepackage[russian]{babel}
\begin{document}

\title{Resonant Transmission of a Light Pulse through a Quantum Well}
\author{L. I. Korovin, I. G. Lang}
\address{A. F. Ioffe Physical-Technical Institute, Russian
Academy of Sciences, 194021 St. Petersburg, Russia}
\author{S. T. Pavlov}
\address{
P.N. Lebedev Physical Institute, Russian Academy of Sciences,
119991 Moscow, Russia; pavlov@sci.lebedev.ru}

\begin{abstract}
Reflectance, transmittance and absorbance of a symmetric light
pulse, the carrying frequency of which is close to the frequency
of interband transitions in a quantum well, are calculated. Energy
levels of the quantum well are assumed discrete, and two closely
located excited levels are taken into account. The theory is
applicable for the quantum wells of arbitrary widths when the size
quantization is preserved. A distinction of refraction indices of
barriers and quantum well is taken into account. In such a case,
some additional reflection from the quantum well borders appears
which changes essentially a shape of the reflected pulse in
comparison to homogeneous medium. The reflection from the borders
disappears at some definite ratios of the carrying frequency of
the stimulating pulse and quantum well width.
\end{abstract}


\maketitle\newpage
\section{Введение}

The optical characteristics (reflectance, transmittance and
absorbance) of a quantum well were calculated \cite{1,2,3,4} for a
symmetric light pulse. A narrow quantum well with one excited
energy level and a homogeneous medium (when the refraction indices
of barriers $\nu_{1}$ and of a quantum well $\nu$ are equal) were
considered \cite{4}. It was assumed in \cite{2} that $\nu =
\nu_{1}$ in a narrow quantum well (QW) with two closely located
excited energy levels. A wide QW (the QW's width is comparable
with a light wave length corresponding to the carrying frequency
of the light pulse) at $\nu = \nu_{1}$ is considered in \cite{4}.
A wide QW with one excited energy level at $\nu\neq\nu_{1}$ is
considered in \cite{3}. It was shown that the heterogeneous medium
influenced essentially the shapes of the reflected and transmitted
pulses. The reflected pulses undergo the greatest changes.

The present work is devoted to an investigation of the time and
spatial dependencies of the reflectance and transmittance of a
symmetrical light pulse going through a QW having a narrow doublet
of the excited energy levels and under condition $\nu\neq\nu_{1}$.
This question is of interest, since the relaxation processes of
the system influence a distortion of the reflected and transmitted
pulses and the correlation of lifetimes of two excited energy
levels. In a homogeneous medium, the reflected pulse depends only
on the resonance with the discrete energy levels of the QW. If
$\nu\neq\nu_{1}$, an additional reflection from the QW borders
appears. An interference of the additional contributions results
in an unconventional dependance of the optical characteristics on
the QW width.

It is assumed that the contributions of the radiative and
nonradiative  relaxation mechanisms may be comparable at low
temperatures, low impurity doping and perfect QW boundaries. It
means that one have to take into account all the orders on the
electron - electromagnetic field interaction
\cite{5,6,7,8,9,10,11}. The estimates \cite{12} show that the
preservation of the size quantized energy levels is possible at
 $kd\geq 1$ ($d$ is the QW width, $k$ is the module of the wave
  vector of the electromagnetic wave corresponding
 to the carrying frequency of the light pulse). In such a case one
 has to take into account a spatial dispersion of waves composing
 the light pulse\cite{12,13,14}.

A system of a semiconductor QW of type I located in the interval
$0\leq z \leq d$ and two semi-infinite barriers is considered. The
system is situated in a constant quantizing magnetic field
directed perpendicularly to the QW plane $xy$ and providing a
discreteness of the energy levels.  A stimulating light pulse
propagates along the $z$ axis from the side of negative values
$z$. The barriers are transparent for the light pulse which is
absorbed in the quantum well to initiate the direct interband
transitions. The intrinsic semiconductor and zero temperature are
assumed.

The final results for two closely spaced energy levels of the
electronic system in a quantum well are obtained. Effect of other
levels on the optical characteristics may be neglected, if the
carrying frequency $\omega_{\,\ell}$ of the light pulse is close
to the frequencies $\omega_{1}$ and  $\omega_{2}$ of the doublet
levels, and other energy levels are fairly distant. It is assumed
that the doublet is situated near the minimum of the conduction
band, the energy levels may be considered in the effective mass
approximation, and the barriers are infinitely high. In
particular, the narrow doublet may be realized by a magnetopolaron
state \cite{14}.

\section{The Fourier-transforms of electric fields of transmitted and reflected pulses}

Let us consider a situation when a symmetric exciting light pulse
of a circular polarization propagates through a single quantum
well along the $z$ axis from the side of negative values of $z$.
Its electric field is as follows
\begin{eqnarray}
\label{eq1}
 \bf{E}_{0}(z,t)~&=&~\bf{e}_{\ell} E_{0}(z,t)~+~ c.c.,\nonumber\\
 E_{0}(z,t)~&=&~E_{0}e^{-i\omega_{\ell}\,p}\Big\{\Theta
(p)e^{-\gamma_{\ell}\,p/2}\nonumber\\&+&[1-\Theta
(p)]e^{\gamma_{\ell} p/2}\Big\}.
\end{eqnarray}
Here $E_{0}$ is the real amplitude, $p=t-\nu_{1}z/c,~
\bf{e}_{\ell}=(\bf{e}_{x}\pm i\bf{e}_{y})/\sqrt{2}$
 is the circular polarization vector,
$\bf{e}_{x}$ и $\bf{e}_{y}$ are the real unite vectors, $\Theta
(p)$ is the Heaviside function, $\omega_{\ell}$ is the carrying
frequency of the light pulse, $ \gamma_{\ell}$ is the pulse
broadening, $c$ is the light velocity in vacuum. The
Fourier-transform of (\ref{eq1}) is
\begin{eqnarray}
\label{eq2} \bf{E}_{0}(z,\omega)=[\bf{e}_{\ell}E_{0}(\omega)+
\bf{e}_{\ell}^{*}E_{0}(-\omega)]e^{ik_{1}z},\nonumber\\
E_{0}(\omega)=\frac{E_{0}~\gamma_{\ell}}{(\omega-\omega_{\ell})^2+
(\gamma_{\ell}/2)^2},~~~~k_{1}=\frac{\nu_{1}\omega}{c}.
 \end{eqnarray}

The electric field at $z\leq 0$ is a sum of the electric fields of
the stimulating and reflected pulses. Its Fourier-transform may be
represented as

$$\bf{E}^{\ell}(z,\omega)=\bf{E}_{0}(z,\omega)+\Delta\bf{E}^{\ell}(z,\omega),$$
where $\Delta\bf{E}^{\ell}(z,\omega)$ is the electric field of the
reflected pulse
\begin{equation}
\label{eq3}\Delta\bf{E}^{\ell}(z,\omega)=\bf{e}_{\ell}\Delta
E^{\ell}(z,\omega)+\bf{e}_{\ell}^{*}\Delta E^{\ell}(z,-\omega).
\end{equation}
The transmitted pulse propagates in the region $z\geq d$. Its
Fourier-transform is
\begin{equation}
\label{eq4}\bf{E}^{r}(z,\omega)=\bf{e}_{\ell}
E^{r}(z,\omega)+\bf{e}_{\ell}^{*} E^{r}(z,-\omega).
\end{equation}

The functions $\Delta E^{\ell}(z,\omega)$ and $E^{r}(z,\omega)$
may be considered as the scalar amplitudes of monochromatic waves
appearing as a result of the exciting  monochromatic wave
propagation through the QW. The similar problem has been solved in
 the case when the interaction of the electronic
system with the electromagnetic wave could not be considered as a
weak perturbation, as well as  at $kd\neq 0$ and $\nu\neq \nu_{1}$
and there were two closely located excited energy levels
\cite{15}.

It has been shown that
\begin{equation}
\label{eq5}\Delta
E^{\ell}(z,\omega)=\mathcal{C}_{R}e^{-ik_{1}z},~~~~z\leq 0,
\end{equation}
\begin{equation}
\label{eq6}E^{r}(z,\omega)=\mathcal{C}_{T}e^{ik_{1}z},~~~z\geq d,
\end{equation}
$\mathcal{C}_{R}$ and $\mathcal{C}_{T}$ determine the amplitudes
of reflected and transmitted waves, respectively. The following
expressions for them are obtained

\begin{eqnarray}
\label{eq7}\mathcal{C}_{R}&=&E_{0}{\rho\over\Delta},\nonumber\\
\mathcal{C}_{T}&=&4E_{0}\zeta
{1+(-1)^{m_{c}+m_{v}}e^{-ikd}N\over\Delta}e^{-ik_{1}d} ,
\end{eqnarray}
\begin{equation}
\label{eq8}\Delta = L -2(\zeta-1)\big
[(\zeta+1)e^{-ikd}+(-1)^{m_{c}+m_{v}}(\zeta-1)\big ]N,
\end{equation}
\begin{eqnarray}
\label{eq9}\rho &=& 2i(\zeta^{2}-1)\sin (kd) + 2\big
[(\zeta^{2}+1)e^{-ikd}\nonumber\\&+&(-1)^{m_{c}+m_{v}}(\zeta^{2}-1)\big
]N.
\end{eqnarray}
In (\ref{eq7}) - (\ref{eq9}) we used the designations
\begin{eqnarray}
\label{eq10}\zeta &=& k/k_{1} = \nu/\nu_{1},~~~k = \nu
\omega/c,\nonumber\\L &=&
(\zeta+1)^{2}e^{-ikd}-(\zeta-1)^{2}e^{ikd},
\end{eqnarray}
$m_{c}(m_{v})$ is the electron (hole) quantum number of the size
quantization. It was assumed that one pair of numbers $m_{c},
m_{v}$ corresponded to two direct interband transitions.

The dependance on the variable $\omega$ is determined by the
function $N$
\begin{equation}
\label{eq11}N =
\frac{-i(-1)^{m_{c}+m_{v}}~(\varepsilon'/2)~\Omega_{0}~e^{ikd}
}{\tilde{\omega}_{1}\tilde{\omega}_{2}+
i(\varepsilon/2)~\Omega_{0}}.
\end{equation}
The designations
\begin{eqnarray}
\label{eq12} \Omega_{0} &=&  \gamma_{r1}\tilde{\omega}_{2} +
\gamma_{r2}\tilde{\omega}_{1},\nonumber\\
\tilde{\omega}_{j}&=&\omega-\omega_{j}+i\gamma_{j}/2,~~~\tilde{\gamma}_{rj}=
\varepsilon'\gamma_{rj},~~~j=1,2.
\end{eqnarray}
are introduced in  (\ref{eq11}). Here $\omega_{j}$ are the
frequencies of interband transitions on the doublet energy levels
, $\gamma_{j}$ and $\gamma_{rj}\sim e^{2}/(\hbar c\,\nu)$ are the
radiative and nonradiative broadenings of the doublet levels
\cite{14}.

The function $N$ contains the complex value
$\varepsilon=\varepsilon'+i\varepsilon''$ \cite{3,13,15}
\begin{equation}
\label{eq13} \varepsilon=\int_{0}^{d}dz\Phi(z)\Bigg\{
\int_{0}^{z}dy e^{ik(z-y)}\Phi(y)+\int_{z}^{d}d
ye^{ik(y-z)}\Phi(y)\Bigg\}.
\end{equation}
It determines the influence of the spatial dispersion on the
radiative broadening $\varepsilon'\gamma_{rj}$ and shift
$\varepsilon''\gamma_{rj}$ of the doublet levels.

Realizing a derivation of (\ref{eq13}) we assumed the Lorentz
force determined by the external magnetic field exceeded the
Coulomb and exchange forces inside of the electron-hole pair,
 and the dependance of the wave function of the electron-hole pair
 may be represented by the factor $\Phi(z)$~\cite{15}.

\section{The time dependance of the electric field of reflected and transmitted light pulses
}

The time dependance of the amplitudes $E^{\ell}(z,\omega)$ and
$E^{r}(z,\omega)$ is determined by well known formulas
\begin{eqnarray}
\label{eq14} \Delta E^{\ell}(z,t) &\equiv& \Delta E^{\ell}(s) =
\frac{1}{2\pi}\int_{-\infty}^{+\infty} d\omega e^{-i\omega\,s}
\Delta E^{\ell}(z,\omega),\nonumber\\ s &=& t + \nu_{1} z/c,
\end{eqnarray}
\begin{eqnarray}\label{eq15}  E^{r}(z,t) &\equiv&  E^{r}(p) =
\frac{1}{2\pi}\int_{-\infty}^{+\infty} d\omega e^{-i\omega\,p}
E^{r}(z,\omega),\nonumber\\p &=& t - \nu_{1} z/c.
\end{eqnarray}

It is seen from (\ref{eq7}) and (\ref{eq8}) that the amplitudes
$\Delta E^{\ell}(z,\omega)$ and $E^{r}(z,\omega)$ have the
identical denominators and, accordingly, identical poles on the
complex plane $\omega$. Having used (\ref{eq2}),
(\ref{eq7})-(\ref{eq9}) and (\ref{eq11}), we obtain the electric
fields in the integral form:
\begin{eqnarray}
\label{eq16} E^{r}(p)=\frac{2\zeta E_{0}\,\gamma_{\ell}\exp
(-ik_{1}d)}{\pi
 L}\int_{-\infty}^{+\infty}d\omega\nonumber\\\times \frac{
U(\omega)\exp(-i\omega p)}{[(\omega - \omega_{\ell})^{\,2} +
(\gamma_{\ell}/2)^{\,2}] (\omega - \Omega_{1})\, (\omega
-\Omega_{2})},
\end{eqnarray}
\begin{eqnarray}
\label{eq17}\Delta E^{\ell}(s)=\frac{E_{0}\,\gamma_{\ell}} {2\pi
L}\int_{-\infty}^{+\infty}d\omega\nonumber\\\times \frac{ V
(\omega)\exp(-i\omega s)}{[(\omega - \omega_{\ell})^{\,2} +
(\gamma_{\ell}/2)^{\,2}] (\omega - \Omega_{1})\, (\omega
-\Omega_{2})},
\end{eqnarray}
where
\begin{equation}
\label{eq18}U(\omega) = \tilde{\omega} _{1}\,\tilde{\omega}_{2} -
\varepsilon'' \,\Omega_{0}/2,
\end{equation}
\begin{equation}
\label{eq19}V(\omega) = a\,\tilde{\omega} _{1}\,\tilde{\omega}_{2}
+ A\,\Omega_{0}/2.
\end{equation}
The designations
\begin{equation}
\label{eq20}a = 2i(\zeta^{2}-1)\sin (kd),
\end{equation}
\begin{equation}
\label{eq21}A = ia\,\varepsilon -
2i(-1)^{m_{c}+m_{v}}[\zeta^{2}+1+(-1)^{m_{c}+m_{v}}(\zeta^{2}-1)]\,\varepsilon'
\end{equation}
are introduced in (\ref{eq19}). The denominators of integrands of
(\ref{eq16}) and (\ref{eq17}) have four identical poles on the
complex plane $\omega$: $\omega = \omega_{\ell}\pm
i\gamma_{\ell}/2,\, \omega = \Omega_{1},\, \omega = \Omega_{2}$.
The pole in the upper semi-plane $\omega = \omega_{\ell} +
i\gamma_{\ell}/2$ defines the time dependance at $p(s)\leq 0$,
three remainder poles are at $p(s)\geq 0$. The poles $\Omega_{1}$
and $\Omega_{2}$ have the view
\begin{eqnarray}
\label{eq22} \Omega_{1(2)}={1\over 2} \bigg
\{\omega_{1}+\omega_{2}-(i/2)\,[\gamma_{1}+\gamma_{2}+f
(\gamma_{r1}+\gamma_{r2})]\nonumber\\\pm \Big[\big
\{\omega_{1}-\omega_{2}-(i/2)\,[\gamma_{1}-\gamma_{2}+
f(\gamma_{r1}-\gamma_{r2})]\big \}^{2}\nonumber\\-
f^{2}\gamma_{r1}\gamma_{r2}\Big]^{1/2}\bigg \}.
\end{eqnarray}

The complex coefficient $f = f_{2} + if_{1}$ enters in
$\Omega_{1}$ and $\Omega_{2}$ as combinations $f\,\gamma_{r1}$ and
$f\,\gamma_{r2}$,
\begin{equation}
\label{eq23}f_{1} = \varepsilon'' -
\frac{(-1)^{m_{c}+m_{v}}\varepsilon'(1-\zeta^{2})\sin
kd}{1+\zeta^{2}+(-1)^{m_{c}+m_{v}}(1-\zeta^{2})\cos kd},
\end{equation}
\begin{equation}
\label{eq24}f_{2} =
\frac{2\varepsilon'\zeta}{1+\zeta^{2}+(-1)^{m_{c}+m_{v}}(1-\zeta^{2})\cos
kd}.
\end{equation}

The mentioned above four poles determine the resonant
contributions  at the integration of (\ref{eq14}) and
(\ref{eq15}). There are a row of poles connected with nulls of the
function $L$, situated on the lower semi-plane $\omega$. However,
these poles are located far away of the real axis,  Однако эти
полюсы расположены далеко от вещественной оси, and their
contributions may be neglected in comparison with the resonant
terms.

The time dependance of the electric fields may be represented as
\begin{eqnarray}
\label{eq25}E^{r}(p) &=& (4\zeta E_{0}/L)\{[1-\Theta
(p)]J_{1}\nonumber\\&+&(J_{2}+J_{3}+J_{4})\Theta (p)\}\exp(-ik_{1}d),\nonumber\\
\Delta E^{\ell}(s) &=& (E_{0}/L)\{[1-\Theta
(s)]K_{1}+(K_{2}\nonumber\\&+&K_{3}+K_{4})\Theta (s)\},
\end{eqnarray}
where
\begin{equation}
\label{eq26}J_{1(2)} = \frac{U(\omega_\ell \pm
i\gamma_\ell/2)\exp[-i(\omega_\ell \pm i\gamma_\ell/2)\,p\,]
}{(\omega_\ell - \Omega_{1} \pm i\gamma_\ell/2)\,(\omega_\ell -
\Omega_{2} \pm i\gamma_\ell/2)},
\end{equation}
\begin{equation}
\label{eq27} J_{3(4)} =
\mp\frac{i\gamma_{\ell}U(\Omega_{1(2)})\exp (-i\Omega_{1(2)}\,p)
}{(\Omega_{1} - \Omega_{2})\,[(\omega_{\ell} -
\Omega_{1(2)})^{\,2} + (\gamma_{\ell}/2)^{\,2}]}
\end{equation}
\begin{equation}
\label{eq28}K_{1(2)} = \frac{ V(\omega_\ell \pm
i\gamma_\ell/2)\exp[-i(\omega_\ell \pm
i\gamma_\ell/2)\,s\,]}{(\omega_\ell - \Omega_{1} \pm
i\gamma_\ell/2)\,(\omega_\ell - \Omega_{2} \pm i\gamma_\ell/2)},
\end{equation}
\begin{equation}
\label{eq29} K_{3(4)} = \mp\frac{i\gamma_{\ell}\,
V(\Omega_{1(2)})\exp (-i\Omega_{1(2)}\,s)}{(\Omega_{1} -
\Omega_{2})\,[(\omega_{\ell} - \Omega_{1(2)})^{\,2} +
(\gamma_{\ell}/2)^{\,2}]}.
\end{equation}

 After the substitutions
$\tilde{\omega}_{1}, \tilde{\omega}_{2}$ and $\Omega_{0}$, the
functions $U(\omega)$ and $V(\omega)$ have the form
\begin{eqnarray}
\label{eq30}&&U(\omega) = (\omega - \omega_{1} +
i\gamma_{1}/2)(\omega - \omega_{2} +
i\gamma_{2}/2)-(\varepsilon''/2)\nonumber\\
&&\times[\gamma_{r1}(\omega - \omega_{2} + i\gamma_{2}/2) +
\gamma_{r2}(\omega - \omega_{1} + i\gamma_{1}/2)],
\end{eqnarray}
\begin{eqnarray}
\label{eq31}&&V(\omega) = a\,(\omega - \omega_{1} +
i\gamma_{1}/2)(\omega - \omega_{2} +
i\gamma_{2}/2) +(A/2)\nonumber\\
&&\times[\gamma_{r1}(\omega - \omega_{2} + i\gamma_{2}/2) +
\gamma_{r2}(\omega - \omega_{1} + i\gamma_{1}/2)].
\end{eqnarray}

Obtaining (\ref{eq22}) для $\Omega_{1}$,  $\Omega_{2}$ we assumed
that $\omega_{\ell}$ was close to the resonant frequencies
$\omega_{1}$ and $\omega_{2}$. Therefore the parameters $k$ and
$k_{1}$ are $k=\nu\omega_{\ell}/c,~~~
k_{1}=\nu_{1}\omega_{\ell}/c$. The analogical approximation was
used in \cite{3}.

\section{Extreme cases}

It follows from (\ref{eq25}) and (\ref{eq31}) that $\Delta
E^{\ell}(s)$ may be represented as a sum of two summands which are
proportional to $a$ and $A$, respectively. The coefficient $a$
equals 0 in two extreme cases: if $\zeta = 1$ (the homogeneous
medium approximation), or $kd = 0$ (a narrow QW). The second
summand equals 0, if $\gamma_{r1} = \gamma_{r2} = 0 $. Hence, one
may conclude that the first summand is stipulated by reflection
from the QW boundaries, and the second by the resonance with the
QW energy levels.

It is seen from (\ref{eq13}), (\ref{eq23}) and(\ref{eq24}) that in
the case $kd = 0~ \varepsilon '' = f_{1} = 0,~ \varepsilon ' = 1,
f_{2} = \zeta$ (if $m_{c} + m_{v}$ is an even number) and $f_{2} =
1/\zeta$~ ($m_{c} + m_{v}$ is an odd number). Thus, in such case
 $\gamma_{r1}$ и $\gamma_{r21}$ are renormalized by the factor
 $\zeta$ (or $\zeta^{-1}$), and in the rest the formulas for the fields coincide
 with the formulas obtained in
 \cite{2}.
If $\zeta = 1,~ kd \neq 0$, (\ref{eq16}) and (\ref{eq17}) pass
into the formulas obtained in \cite{4}.

Below a calculation is being provided for the case
$$\gamma_{r1} = \gamma_{r2} = \gamma_{r},~~\gamma_{1} = \gamma_{2} = \gamma.$$
Such equality is possible, for instance, if the doublet is formed
by a magnetopolaron when the cyclotron  frequency is equal to the
longitudinal optical phonon frequency \cite{14}. In such limiting
case  $\Omega_{1(2)}$  (\ref{eq22}) equal:
\begin{equation}
\label{eq32}\Omega_{01(2)}={1\over 2}\bigg[\omega_{1} + \omega_{2}
- i(\gamma + f\gamma_{r}) \pm \sqrt{(\omega_{1} - \omega_{2})^{2}
- f^{2}\gamma_{r}^{2}}~\bigg].
\end{equation}

For an homogeneous medium $(\zeta = 1)$ it follows from
(\ref{eq23}) and (\ref{eq24}) that $f = \varepsilon$ and
$\Omega_{01(2)}$ coincides with the analogical formulas from
\cite{4}. Thus, a transition to a heterogeneous medium leads to
the replacement in the resonant denominators in formulas
(\ref{eq26}) - (\ref{eq29}) the function $\varepsilon$ by the
coefficient $f$ which influences now the shift  $f_{1}$ and the
radiative broadening  $f_{2}$ of the doublet energy levels. The
analogical situation, as it is seen from (\ref{eq22}), takes place
and in the general case $\gamma_{r1} \neq \gamma_{r2}, \gamma_{r}
\neq \gamma_{r}$.

For the sake of convenience of numerical calculations, let us go
to the new variables
\begin{eqnarray}
\label{eq33}\Omega &=& \omega_{\ell} - \omega_{1},~~\Delta\omega =
\omega_{1} - \omega_{2},~~\beta_{01} = \Omega_{01} -
\omega_{1},\nonumber\\ \beta_{02} &=&\Omega_{02} - \omega_{1},
\end{eqnarray}
(with the help of these variables the number of the independent
parameters in expressions for optical characteristics diminishes
on one), and to introduce also the resonant frequency
\begin{equation}
\label{eq34}\Omega_{res} \equiv Re~\beta_{01} ={1\over 2} \bigg
[-\Delta\omega + f_{1}\gamma_{r} +Re~\sqrt{\Delta\omega^{2} -
f^2\gamma_{r}^2}~\bigg ],
\end{equation}
corresponding to the interband resonant transition with the
frequency $\omega_{1}$, renormalized by the radiative shift.

In this extreme case, the electric fields take the form
\begin{eqnarray}
\label{eq35}E^{r}(p) &=& (4\zeta E_{0}/L)e^{-i(k_1
d+\omega_{\ell}p)}\{[1-\Theta
(p)\,]J_{01}\nonumber\\&+&(J_{02}+J_{03}+J_{04})\Theta (p)\},\nonumber\\
\Delta E^{\ell}(s) &=& (4 E_{0}/L)e^{-i\omega_{\ell}s}\{[1-\Theta
(s)]K_{01}\nonumber\\&+&(K_{02}+K_{03}+K_{04})\Theta (s)\},
\end{eqnarray}
\begin{equation}
\label{eq36}J_{01(2)} = \frac{\exp(\pm
\gamma_\ell\,p/2)~U_{0}(\Omega \pm i\gamma_\ell/2)}{(\Omega -
\beta_{01} \pm i\gamma_\ell/2)\,(\Omega - \beta_{02} \pm
i\gamma_\ell/2)},
\end{equation}
\begin{equation}
\label{eq37} J_{03(4)} = \mp\frac{i\gamma_{\ell}\exp [i(\Omega -
\beta_{01(2)})\,p\,]~U_{0}(\Omega_{01(2)})}{\sqrt{\Delta\omega^{2}
- f^{2}\gamma_{r}^{2}}~[(\Omega - \beta_{01(2)})^{\,2} +
(\gamma_{\ell}/2)^{\,2}\,]\,},
\end{equation}
\begin{equation}
\label{eq38}K_{01(2)} = \frac{\exp(\pm \gamma_\ell\,
s/2)~V_{0}(\Omega \pm i\gamma_\ell/2)}{(\Omega - \beta_{01} \pm
i\gamma_\ell/2)\,(\Omega - \beta_{02} \pm i\gamma_\ell/2)},
\end{equation}
\begin{equation}
\label{eq39} K_{03(4)} = \mp\frac{i\gamma_{\ell}\exp [i(\Omega -
\beta_{01(2)})\,s\,]~V_{0}(\Omega_{01(2)})}{\sqrt{\Delta\omega^{2}
- f^{2}\gamma_{r}^{2}}~[(\Omega - \beta_{01(2)})^{2} +
(\gamma_{\ell}/2)^{2}\,]}.
\end{equation}
Instead of (\ref{eq30}) and (\ref{eq31}) the functions
\begin{eqnarray}
\label{eq40}U_{0}(\omega) &=& (\omega + i\gamma/2)(2\omega +
\Delta\omega +
i\gamma)\nonumber\\
&-&(\varepsilon''\gamma_{r}/2)(2\omega + \Delta\omega +i\gamma),
\end{eqnarray}
\begin{eqnarray}
\label{eq41}V_{0}(\omega) &=& a(\omega + i\gamma/2)\,(\omega
+\Delta\omega +
i\gamma/2)\nonumber\\
&+&(A~\gamma_{r}/2)(2\omega + \Delta\omega +i\gamma),
\end{eqnarray}
appear, where $a$ and $A$ are determined in (\ref{eq20}),
(\ref{eq21}).

It is of interest a case when  $\gamma_{r1} = \gamma_{r2} =
0$,i.e., the interaction of the electromagnetic waves with the
electronic system is absent. It may happen for the frequencies
faraway from the resonance when the absorption is neglected. In
such a case it follows from (\ref{eq25}) - (\ref{eq31}) that
\begin{equation}
\label{eq42}E^{r}(p) = (4\zeta/L)E_{0}(z,t),~~~\Delta E^{\ell}(s)
= (a/L)E_{0}(z,t),
\end{equation}
where $E_{0}(z,t)$ is the scalar amplitude of the field of the
exciting pulse (\ref{eq1}). Thus, for an transparent QW the
electric fields of transmitted and reflected pulses are
proportional to the field of the exciting pulse. The corresponding
coefficients depend on $\zeta$ and $kd$.

\section{Reflection, transition and absorption of the exciting pulse}

 The energy flux $\bf{S}(p)$, corresponding to the electric field of the exciting pulse, is defined as \cite{13}
\begin{equation} \label{eq43}\bf{S}(p) =
(\bf{e}_{z}/4\pi)(c\,\nu_{1})(\bf{E}_{0}(z,t))^{2} =
\bf{e}_{z}S_{0}\mathcal{P}(p) ,
\end{equation}
where $S_{0} = c\nu_{1}E_{0}^{2}/(2\pi)$, $\bf{e}_{z}$ is the
unite vector along the axes $z$. The dimensionless function
 $\mathcal{P}(p)$ determines the spatial and time dependencies of the energy flux of the exciting pulse,
\begin{equation} \label{eq44} \mathcal{P}(p)={(\bf{E}_{0}(z,t))^{2}\over S_{0}}=\Theta
(p)e^{-\gamma_{\ell}p}+[1-\Theta (p)]e^{\gamma_{\ell}p}.
\end{equation}
Transmitted (on the right of the QW) and reflected (on the left of
the QW) fluxes are determined as
\begin{eqnarray}
\label{eq45} \bf{S}^{r}(z,t) =
 {\bf{e}_{z}\over 4\pi}c\nu_{1}(\bf{E}^{r}(z,t))^{2} =
\bf{e}_{z}S_{0}\mathcal{T}(p),\nonumber\\
\bf{S}^{\ell}(z,t) =
 -{\bf{e}_{z}\over 4\pi}c\nu_{1}(\Delta\bf{E}^{\ell}(z,t))^{2}=-\bf{e}_{z}S_{0}{\mathcal{R}}(s),
\end{eqnarray}
dimensionless functions  $\mathcal{T}(p)$ and $\mathcal{R}(s)$
determine the parts of the transmitted and reflected energy of the
stimulating pulse. The energy part $\mathcal{A}(t)$, concentrated
inside of the QW, is being absorbed or irradiated again. It is
defined by the obvious expression
\begin{equation} \label{eq46} \mathcal{A}(p) = \mathcal{P}(p) - \mathcal{R}(p)
-\mathcal{T}(p)
\end{equation}
(since for reflection $z\leq 0$, the variable in $\mathcal{R}$ is
$s=t-|z|\nu/c$).

In the limiting case $\gamma_{r1} = \gamma_{r2} =
\gamma_{r}~~\mathcal{A}= 0,\,
 \mathcal{T} = |4\zeta/L|^{2}\mathcal{P}(t)$, $\mathcal{R} = |a/L|^2 \mathcal{P}(t)$, where $P(t)$
corresponds to the stimulating pulse. Taking into account
(\ref{eq10}) we obtain
\begin{eqnarray} \label{eq47}&&\mathcal{R}(t) =  \frac{(\zeta^{2}-1)^{2}\sin ^{2}(kd)}
{4\zeta^{2}\cos^{2}(kd) + (\zeta^{2}+1)^{2}\sin^{2}(kd)}\mathcal{P}(t),\nonumber\\
&& \mathcal{T}(t) =  \frac{4\zeta^2}{4\zeta^{2}\cos^{2}(kd) +
(\zeta^{2}+1)^{2}\sin^{2}(kd)}\mathcal{P}(t).
\end{eqnarray}
Thus, in this limiting case $\mathcal{R}$ is the reflectance of a
transparent plate and equals 0 at $kd = 0, \pi, 2\pi...$.

For the calculations one has to define the function $\Phi (z)$,
from the definition of $\varepsilon$ (\ref{eq13}). $\Phi (z)$ was
chosen as
$$\Phi (z) = (2/d)\sin(\pi m_{c} z/d)\sin(\pi m_{v} z/d),~~~~0\leq z \leq d$$
and $\Phi (z) = 0$ in the barriers, what corresponds to the free
electron-hole pair. In the case of chosen above $\Phi
(z)~\varepsilon'$ and $\varepsilon''$, according to (\ref{eq13}),
are equal
\begin{eqnarray} \label{eq48}&& \varepsilon' = 2\mathcal{B}^{2}[1-
(-1)^{m_{c}+m_{v}} \cos (kd)], \nonumber\\&& \varepsilon'' =
2\mathcal{B}\Big (
\frac{(1+\delta_{m_{c}+m_{v}})(m_{c}+m_{v})^{2}+(m_{c}-m_{v})^2}{8m_{c}m_{v}}\nonumber\\&&
-(-1)^{m_{c}+m_{v}}\mathcal{B}\sin kd  -
\frac{(2+\delta_{m_{c}+m_{v}})(kd)^{2}}{8\pi^{2}m_{c}m_{v}}\Big ),
\end{eqnarray}
$$ \mathcal{B} = \frac{4\pi^{2}m_{c}m_{v}kd}{[\pi^{2}(m_{c}+m_{v})^{2}-
(kd)^{2}]\, [(kd)^{2}-\pi^{2}(m_{c}-m_{v})^{2}]}$$
\section{Calculation results and discussion}

The poles $\Omega_{1}$ and $\Omega_{2}$ (\ref{eq22}) determine the
resonant frequencies  $Re~\Omega_{1,2}$ and shift
$Im~\Omega_{1,2}$ of the doublet levels. In the limiting case of a
homogeneous medium and a narrow QW  $\Omega_{1,2}$ coincide with
values $(\Omega - iG/2)_{1(2)}$ obtained in \cite{2}. The
broadenings $\gamma_{ri}$ of the doublet levels enter into
electric fields of the pulse as $\varepsilon'\,\gamma_{ri}$ and
$\varepsilon''\gamma_{ri}$. In the cases of the monochromatic and
pulse irradiation in a homogeneous medium
$\varepsilon'\,\gamma_{ri}$ and $\varepsilon''\,\gamma_{ri}$
determine the broadening and shift of the doublet levels \cite{13,
15, 4}. In the considered above general case of the pulse
irradiation and heterogeneous medium,  $\Omega_{1}$ and
$\Omega_{2}$ contain $f_{1}\gamma_{ri}$ and $f_{2}\gamma_{ri}$
which depend on parameters $kd$ and $\zeta$ and on
$\varepsilon'\,\gamma_{ri}$ and $\varepsilon''\,\gamma_{ri}$.
Thus, the view of functions $f_{2}$ and $f_{1}$ influence the
radiative shift and broadening of the energy levels.

If $\zeta = 1, kd \neq 0$, then $f_{1} = \varepsilon'', f_{2} =
\varepsilon'$. In the case $kd = 0$ the level shift is absent, and
the broadening is determined by values $\zeta \gamma_{ri}$. Since
at $kd \neq 0~~\gamma_{ri} \sim \nu^{-1}$, then $\zeta \gamma_{ri}
\sim \nu_{1}^{-1}$ and is some broadening at zero width of a QW
\cite{15}.
 \begin{figure}
 \includegraphics [] {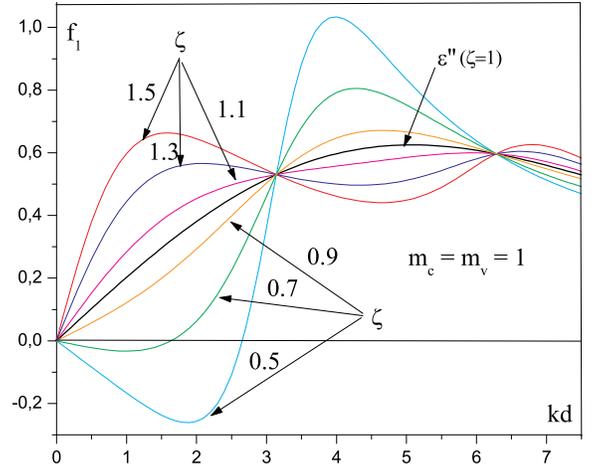} 
 \caption[*]{\label{Fig.1.eps} $f_{1}$ as function of $kd$
(\ref{eq23}))
 at different values $\zeta = \nu/\nu_{1}$.}
 \end{figure}
 \begin{figure}
 \includegraphics [] {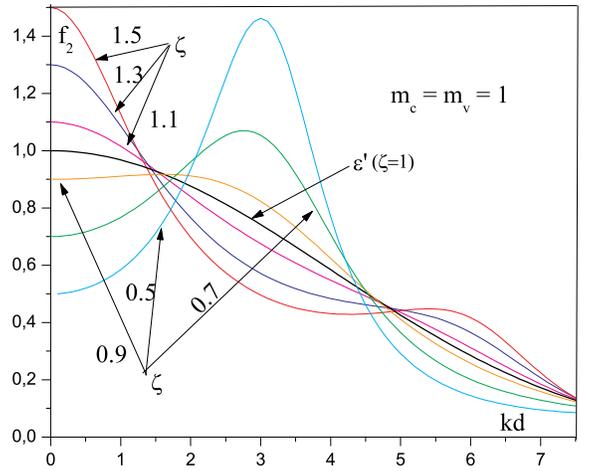} 
 \caption[*]{\label{Fig.2.eps} $f_{2}$ as function of $kd$
(\ref{eq24})
 at different values $\zeta = \nu/\nu_{1}$.}
 \end{figure}

In Fig.1  $f_{1}$ from  (\ref{eq23}) (which is connected with the
shift of the doublet energy levels) is represented as a function
of  $kd$ for different values $\zeta$. It is seen that $f_{1} =
\varepsilon''$ in the points $kd =0, \pi, 2\pi... $, i.e.,the
influence of the heterogeneous medium on the level shift
disappears. $f_{2}$ ((\ref{eq24})) as a function of $kd$ is
represented in Fig.2.  The largest deviation $f_{2}$ from
$\varepsilon'$ takes place in the points $kd = 0, \pi, 2\pi ...
f_{2}$, since in these points $f_{2} = \zeta$.

The functions $\mathcal{R}, \mathcal{A}$ and $\mathcal{T}$ were
calculated for $\gamma_{r1} = \gamma_{r2} =
\gamma_{r}$,~~$\gamma_{1} = \gamma_{2} = \gamma$ according to
(\ref{eq35}). It was assumed the direct interband transition with
the quantum numbers $m_{c} = m_{v} = 1$. It is shown in figures
changes in time of the optical characteristics of a QW at passing
of a light pulse for different values of parameters $kd$ and
$\zeta = \nu/\nu_{1}$. Since the functions $\mathcal{R}$ and
$\mathcal{T}$ are the homogeneous functions of the broadenings and
of frequencies $\omega_{1}, \omega_{2}, \omega_{\ell}$, then the
choice of the measuring units is unconditioned. All the values are
expressed in $eV$ for the sake of certainty. All the curves
$\mathcal{R, \mathcal{T }}$ and $\mathcal{A}$ are obtained for the
case $\Omega = \Omega_{res}$, where ~ $\Omega$ and $\Omega_{res}$
are determined by the formulas (\ref{eq33}), (\ref{eq34}).
 \begin{figure}
 \includegraphics [] {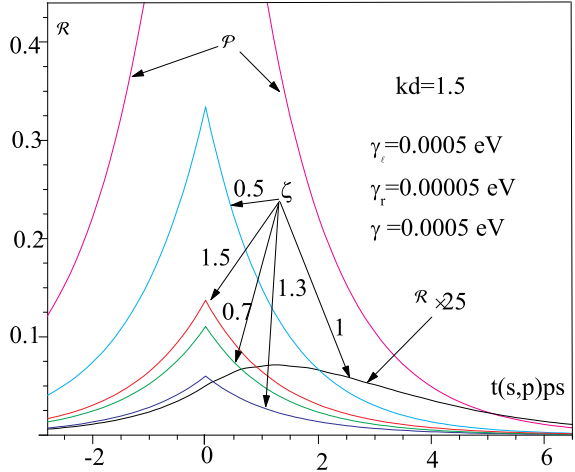} 
 \caption[*]{\label{Fig.3.eps}The reflectance $\mathcal{R}$ as a
 function of time for a narrow (in comparison to
 $\Delta\omega$) stimulating pulse
$(\Delta\omega = 0.0065~ eV)$ and small radiative broadening
$\gamma_{r}\ll\gamma, \gamma_{\ell}$.}
 \end{figure}

The reflectance $\mathcal{R}$ for an narrow in comparison with
$\Delta\omega$ stimulating pulse and a small radiative broadening
$(\gamma_{r}\ll \gamma, \gamma_{\ell})$ is represented in Fig.3.
All the curves correspond to the value kd=1.5. In such a case, as
it follows from Fig.1,2, the radiative broadening is close
$\varepsilon' \gamma_{r}$ and weakly dependant on $\zeta$. On the
other hand, the radiative shift of energy levels
$f_{1}=\varepsilon''-(1-\zeta^{2})/(1+ \zeta^{2})$ depends
substantially on the parameter $\zeta$. It is seen that the
reflectance changes radically for an homogeneous medium
$(\zeta=1)$ in comparison with a heterogeneous medium
  $(\zeta\neq 1)$. For instance, at $\zeta = 1.3$ the reflectance in maximum
  increases 25 times in comparison with the case of
  $\zeta = 1$, at $\zeta = 0.7$ - 50 times. The sharp increasing of $\mathcal{R}$
 is a manifestation of light reflection from the QW borders which
 equals 0 at $\zeta = 1$.
 \begin{figure}
 \includegraphics [] {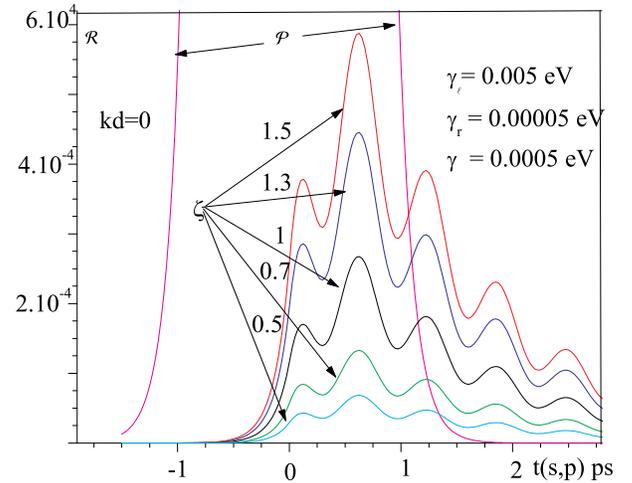} 
 \caption[*]{\label{Fig.4.eps}The time dependance of the reflectance $\mathcal{R}$
for an exciting pulse of a middle duration $(\Delta\omega = 0.0065
eV, \gamma_{\ell} = 0.005 eV)$.  The reflection from the QW
borders is absent $(kd = 0)$. }
 \end{figure}
 \begin{figure}
 \includegraphics [] {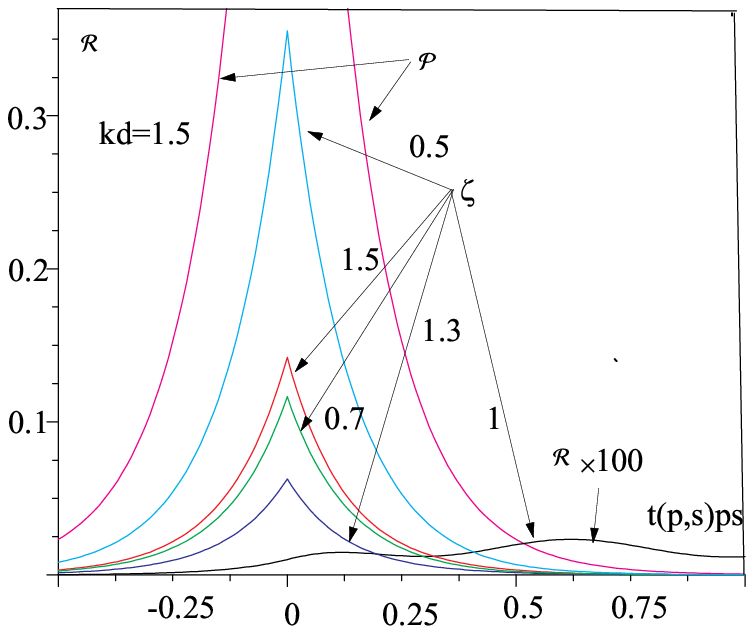} 
 \caption[*]{\label{Fig.5.eps}The time dependance of the reflectance $\mathcal{R}$
for an exciting pulse of a middle duration $(\Delta\omega = 0.0065
eV, \gamma_{\ell} = 0.005 eV)$.  The reflection from the QW
borders is maximal $(kd = 1.5)$.}
 \end{figure}
 \begin{figure}
 \includegraphics [] {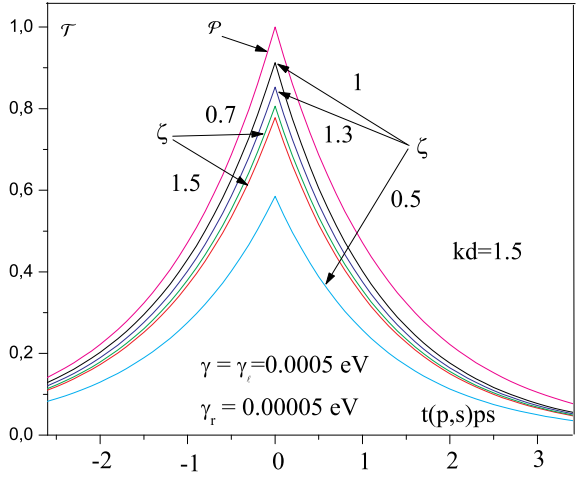} 
 \caption[*]{\label{Fig.6.eps}The transmittance $\mathcal{T}$ in the case of maximal reflection from the QW borders
  $(kd = 1.5)$. The parameters are the same as in Fig.4, 5.}
 \end{figure}

An analogical situation is shown in Fig.4,5, which  demonstrate a
case of a light pulse of a middle duration when
 $\gamma_{\ell}\cong\Delta\omega$ and $\gamma_{r}\ll\gamma\ll\gamma_{\ell}$. In Fig.4 $(kd =
 0)$) a reflection from boundaries is absent and a dependence of the reflectance on $\zeta$
 is determined only by the parameter $\zeta\gamma_{r}$. In Fig.5 $(kd = 1.5)$, where the reflection from the
 boundaries is essential, there is a sharp increasing of reflection
 in comparison with the case of $kd =
 0$: the ratio of  $\mathcal{R}(\zeta \neq 0)/\mathcal{R}(\zeta = 0)$ in maximum increases in 200 times
  $(\gamma = 1.3)$ and in 1100 times $(\zeta = 0.5)$ раз.

 \begin{figure}
 \includegraphics [] {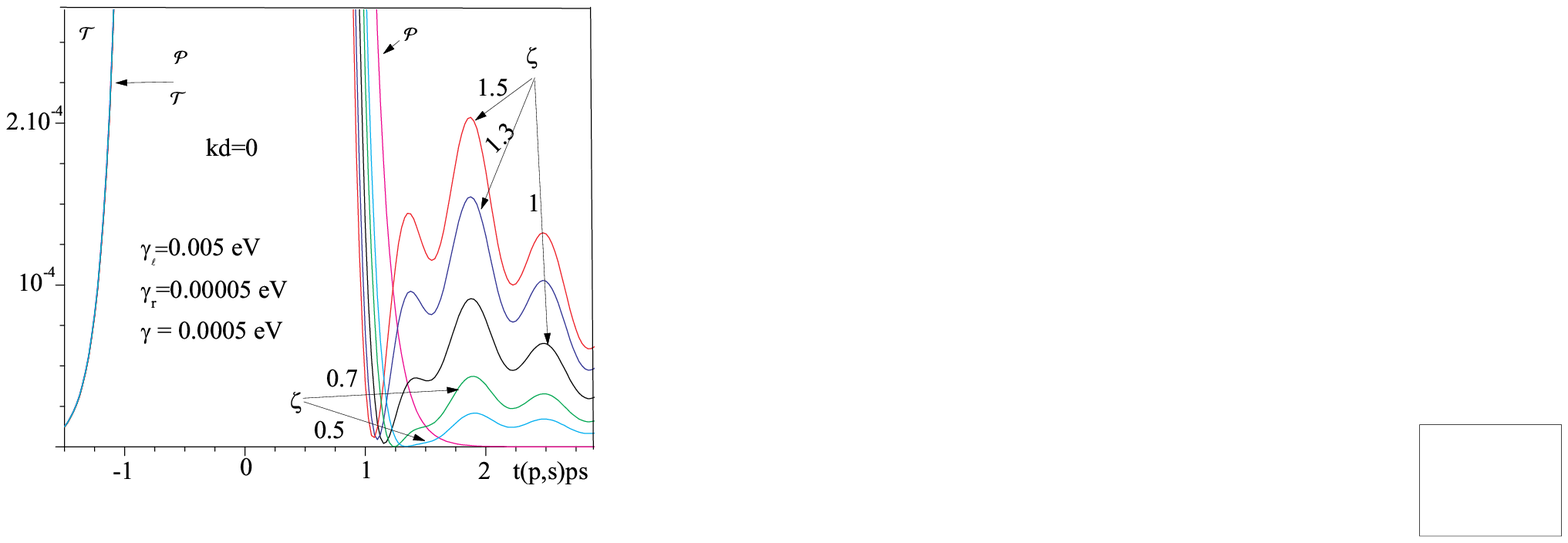} 
 \caption[*]{\label{Fig.7.eps}The time dependance of the transmittance $\mathcal{T}$
 in the case of narrow quantum wells $(kd =
0)$, when the reflection from the QW borders is absent.
$\Delta\omega = 0.0065 eV$.}
 \end{figure}
 \begin{figure}
 \includegraphics [] {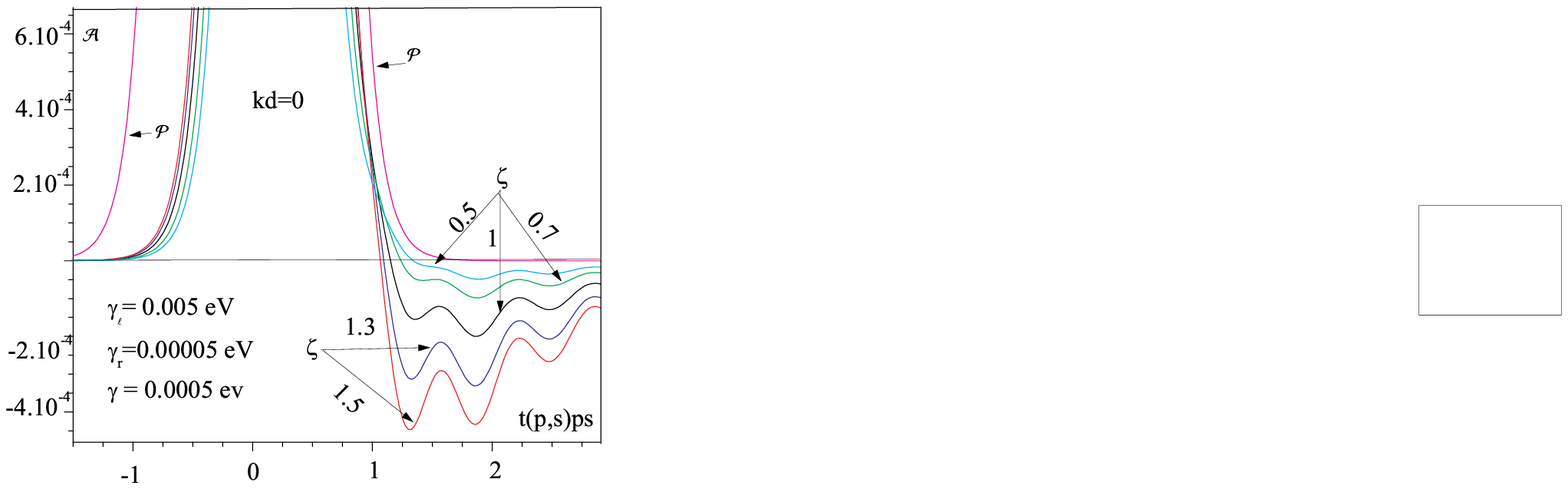} 
 \caption[*]{\label{Fig.8.eps}The time dependance of the absorbance $\mathcal{A}$ in the
  case of narrow quantum wells $(kd = 0)$,
 when the reflection from the QW borders is absent. $\Delta\omega = 0.0065
eV$.}
 \end{figure}
  Figs.6, 7, 8 demonstrate an influence of a heterogeneous medium
   on the transmittance $\mathcal{T}$ and on the energy part
   accumulated by the QW in the resonant transitions
  $\mathcal{A}$. In Fig.6 the parameters $\gamma_{\ell}, \gamma $,  $\gamma_{r}$
 are the same as in Fig.3, and in Fig.7, 8 they are the same as in Fig.4, 5.
In Fig.6 the curves $\mathcal{T}(p)$ are represented for $kd =
1.5$, when a boundary reflection is largest one.
 At $kd = 0$ and $kd = \pi$ that reflection disappears and an influence of
  $\zeta$ influences weakly on $\mathcal{T}(p)$. In Fig.7, 8 the curves $\mathcal{T}$
  and
 $\mathcal{A}$ are calculated for $kd = 0$. In such a case, the
 dependencies
  $\mathcal{T}(p)$ and $\mathcal{A}$ on $\zeta$ are stipulated only by the parameter
   $\zeta\gamma_{r}$.  There appears a generation
 (a negative absorption) after pulse passing  (Fig.8). The generation is stipulated
 by that that the electronic system does not absorb or radiate completely  the energy, accumulated in resonant
 transitions during the time the pulse passes through the QW.

 It is follows from the results represented above that taking into
 account the differences of the refraction indexes of the QW and
 barriers influences noticeably the reflectance. It is most strong
 when a reflection, stipulated by the resonant transitions in the
 QW, is a small value. The reflection from the QW boundaries
 depends on
$kd$ and disappears at $kd = p\pi$,
 where $p$ is an integer. In these points the system QW - barriers
 looks like a homogeneous medium. A distinction from a homogeneous medium is only in a substitution of
 the radiative broadening $\epsilon'\gamma_{r}$ by
 $\epsilon'\zeta\gamma_{r}$.

 The QW optical properties are like the optical properties
 of a plate with the parallel borders inserted into a medium with different refraction index.
 Indeed, if an absorption is small
  (for instance, if a carrying frequency is far away from the resonant frequencies of absorption
 ), then $\mathcal{R}$ and $\mathcal{T}$
 are described by (\ref{eq47}), which is just fair for such a plate under a pulse irradiation
 . A sharp decrease of reflection in the points  $kd =
 p\pi$ takes place also nearby the resonant absorption frequencies.

\end{document}